\documentclass[aps,prd,groupedaddress,showpacs,showkeys]{revtex4}
\usepackage{amssymb}
\usepackage{epsfig}

\newcommand{\eq}{\begin{eqnarray}}
\newcommand{\en}{\end{eqnarray}}

\begin{document}

\title{Reply to comment on\\
``Two-photon decay of the sigma meson''}
\author{ Francesco Giacosa$^1$, Thomas Gutsche$^2$, 
Valery E. Lyubovitskij$^2 $\footnote{
On leave of absence from the Department of Physics, Tomsk State University,
634050 Tomsk, Russia} \vspace*{1.2\baselineskip}}
\affiliation{$^1$ Institut f\"ur Theoretische Physik, Johann Wolfgang
Goethe-Universit\"at, Max von Laue-Strasse 1, 60438 Frankfurt, Germany 
\vspace*{0.3\baselineskip}\\ 
$^2$ Institut f\"ur Theoretische Physik, Universit\"at T\"ubingen, Kepler
Center for Astro and Particle Physics, Auf der Morgenstelle 14, D-72076
T\"ubingen, Germany \vspace*{0.3\baselineskip}\\
}

\begin{abstract}
We reply to the preceding Comment by E.~van Beveren et al., arXiv:0811.2589
[hep-ph].
\end{abstract}

\pacs{12.39.Ki,13.25.Jx,13.30.Eg,13.40.Hq}
\keywords{$\sigma$-meson, relativistic quark model, electromagnetic decay}
\maketitle

In Ref.~\cite{our} we evaluated the radiative decay width $\sigma
/f_{0}(600)\rightarrow \gamma \gamma $ under the assumption that $\sigma $
is described as a quark-antiquark state with flavor configuration $\overline{%
n}n\equiv (\overline{u}u+\overline{d}d)/\sqrt{2}$. For this purpose we
utilized both local and non-local interaction Lagrangians describing the
coupling of constituent quarks to the scalar field and which also allows a
consistent, gauge-invariant inclusion of the electromagnetic interaction.
Since a non-local model description of a quark-antiquark bound state is in
our view unavoidable, we demonstrated in a pure quarkonium description that $%
\Gamma _{\sigma \rightarrow \gamma \gamma }<1$ keV for a mass of $M_{\sigma
}\lesssim 800$ MeV. In Ref.~\cite{our} we also stress that a further
inclusion of meson loops will enhance the value of $\Gamma _{\sigma
\rightarrow \gamma \gamma }$, but an explicit calculation was not performed.
Our main conclusion in Ref.~\cite{our} was therefore that a dominant or pure
quarkonia interpretation of the $\sigma $ does not allow for a full
explanation of currently available data~\cite{pdg} on $\Gamma _{\sigma
\rightarrow \gamma \gamma }$.

The preceding Comment by E. van Beveren et al.~\cite%
{comment} claims that in ~\cite{our} we were mistaken on essentially three
points on which we briefly elaborate in the following.

\bigskip

\emph{Evaluation of the quark triangle diagram:} In Ref.~\cite{our} we
stress that an accurate evaluation of the quark triangle diagram for $\sigma
\rightarrow \gamma \gamma $ generates a term which in general causes
destructive interference when compared to the corresponding $\pi
^{0}\rightarrow \gamma \gamma $ amplitude. This additional term vanishes
under the peculiar condition $M_{\sigma }=2m_{q}$, where $m_{q}$ is the
constituent quark mass. The original citation in Ref.~\cite{our} of
the authors of preceding Comment with respect to this technical issue
referred to this peculiarity. The analytical results both deduced in Ref.~%
\cite{our} and Ref.~\cite{comment} for $\Gamma _{\sigma \rightarrow \gamma
\gamma }$ now completely agree in the case of a local Lagrangian formulation.

\bigskip

\emph{Nonlocal description of quark-antiquark bound states:} To further
illustrate the need for a nonlocal Lagrangian formulation we first refer to
the Nambu Jona-Lasinio (NJL) model, which originally is given in local form
(see e.g. Refs.~\cite{hk,klevansky,Dorokhov:2003ib}). Regularization of loop
integrals requires the introduction of a sharp cutoff $\Lambda $ or a cutoff
function. Independent of the precise form of the cutoff function the
important point is that the cutoff $\Lambda \sim 1$ GeV has a well-defined
physical meaning: it is related to the nonperturbative nature of the
underlying and fundamental theory of quarks and gluons, QCD, and it sets the
corresponding low-energy scale. Note however that the cutoff $\Lambda $, together with
the precise form of the cutoff procedure is not included in the original NJL
Lagrangian. Once a physical cutoff of the order $\Lambda \sim 1$ GeV has
been introduced in an effective theory, it should be consistently included
in all diagrams, including those that are ultraviolet (UV) convergent. A
simple way to introduce this cutoff function already at the level of the
starting Lagrangian is to render it nonlocal. This is explicitly done, for
instance, in Ref.~\cite{efimov1,efimov2,anikin,volkov}. On a quantitative
level NJL models with a proper introduction of the regularization procedure
deliver an upper bound with $\Gamma _{\sigma \rightarrow \gamma \gamma }<1$
keV in a pure quarkonium interpretation \cite{volkov2}. This finding is
consistent with the
results of Ref.~\cite{our}, although explicit numbers will depend on
dynamical details and for example the explicit values for the $\sigma $ and
constituent quark masses.

Even on more general grounds, the QCD Bethe-Salpeter approach or QCD motived
quark models based on bosonization of the QCD generating functional (for a
review see~\cite{alkofer} and~\cite{efimov1,efimov2}) show that a nonlocal
interaction of a meson with its constituents -the quarks- naturally emerges
out of quark-gluon-dynamics. One might argue about the precise form of
vertex functions and quark propagators, but the very fact that a nonlocal
interaction arises seems undisputable.

We therefore still argue that a non-local description of quark-antiquark
bound states with a typical intrinsic scale of about 1 GeV will result in
values for $\Gamma_{\sigma\rightarrow\gamma\gamma}$ below 1 keV, with
explicit quantitative numbers depending on dynamical details and, trivially,
on the mass of the $\sigma$. Please note that most of the analyses now agree
on a pole position of the $\sigma$ near (500 - i 250) MeV (see also note on
scalar mesons~\cite{pdg}).

\bigskip

\emph{Meson-loops:} Because of the large width of the $\sigma $ the coupling
to $\pi \pi $ and $K\bar{K}$ followed by final state interaction will have a
strong impact on the radiative decay width. In the Comment~\cite{comment}
the authors deduce a net effect due to meson loops of about 40\% of their
total two-gamma width $\Gamma _{\sigma \rightarrow \gamma \gamma }\approx 3.5
$ keV. In Ref.~\cite{volkov2} meson loops also contribute by about 50\% but
resulting only in $\Gamma _{\sigma \rightarrow \gamma \gamma }\approx 1.03$
keV in total. A recent model dependent analysis of $\pi \pi $ and $\gamma
\gamma $ scattering data~\cite{Mennessier:2008kk} deduces a total 2$\gamma $
decay width of $\Gamma _{\sigma \rightarrow \gamma \gamma }^{\mathrm{tot}%
}\approx (3.9\pm 0.6)$ keV, where the bulk part can be explained by
rescattering. The direct or bare $\sigma $-pole contribution results in only 
$\Gamma _{\sigma \rightarrow \gamma \gamma }^{\mathrm{dir}}\approx (0.13\pm
0.05)$ keV, actually in line with our results of Ref.~{\cite{our}}. Again, a
model-independent estimate of meson loop contributions to the 2$\gamma $
decay width of the $\sigma $ seems presently not available. An analysis by
Pennington \cite{pennington} confirms the large value for the $\gamma \gamma 
$ decay width with $\Gamma _{\sigma \rightarrow \gamma \gamma }=(4.1\pm 0.3)$
keV, although Oller et al.~\cite{oller} or Bernabeu et al.~\cite{bernabeu}
deduce in their analyses smaller values of $1.8\pm 0.4$ keV and $1.2\pm 0.4$
keV, respectively.

If a large $\gamma\gamma$ decay width of the $f_{0}(600)$ will be confirmed
in future, our theoretical analysis shows that this result cannot be explained
by the quark-loop contribution alone. Then we have two options: (i)
Discard a dominant quark-antiquark interpretation of the $f_{0}(600),$ in
agreement with many recent works \cite{klempt}; (ii) Argue that the meson
loops generate the -by far- dominant contribution. In this case, however, it
will be rather difficult to extract precise information about the nature of
scalar states from $\gamma\gamma$ decays.

\begin{acknowledgments}
This work was supported by the DFG under Contract No. FA67/31-1, No.
FA67/31-2, and No. GRK683. This research is also part of the European
Community-Research Infrastructure Integrating Activity "Study of Strongly
Interacting Matter" (HadronPhysics2, Grant Agreement No. 227431) and
President grant of Russia "Scientific Schools" No. 871.2008.2.
\end{acknowledgments}

\end{document}